\documentclass[aps,pra,showpacs,twocolumn,superscriptaddress,showpacs,groupedaddress,nofootinbib]{revtex4-1}

\usepackage{graphicx}% Include figure files
\usepackage{dcolumn}% Align table columns on decimal point
\usepackage{bm}% bold math

\usepackage{amssymb}   % for math
\usepackage{amsmath,amssymb,amsthm,mathrsfs,amsfonts,dsfont} 

\usepackage{physics}   % physics package

\usepackage[colorlinks=true,citecolor=magenta, linkcolor=blue]{hyperref}
\usepackage[dvipsnames]{xcolor}

\newcommand{\p}{\partial}
\newcommand{\aq}{\begin{eqnarray}}
\newcommand{\qa}{\end{eqnarray}}

\begin{document}

\title{Obtaining eigenvalues and solutions for a class of differential equations using Casimir operator}

\author{K. Maharana}
\email{karmadev@iopb.res.in}
%\thanks{}
\affiliation{Physics Department, Utkal University, Bhubaneswar 751004, 
  India.}

\date{\today}

\begin{abstract}
{Using the group theoretic method of spectrum generating algebras
a class of   differential equations is obtained whose eigenvalues
are calculated without explicitly 
solving the equations. Solutions can be easily obtained by
group theoretic methods  for a certain type of potentials.}
\end{abstract}

\pacs{02.20.-a, 02.30.Gp, 02.20.Hq}  

\maketitle

A basic objective  of quantum mechanics is to find the eigenfunctions
and the eigenvalues for a given potential. The common practice, started by 
 Schr\"odinger, is to first set up the  Hamiltonian and solve  the differential equation to 
 obtain the eigenfunctions and the eigenvalues.  However,
 historically, the action angle variables of Hamilton-Jacobi theory determined the 
 periods without giving the solutions with time development\cite{Jammer}.  
Pauli obtained the energy eigenvalues of the hydrogen atom using the
group theoretic method that also clearly demonstrates the extra symmetries that arise
in the form of Runge-Lenz vectors\cite{Pauli}. The method uses
algebriac techniques to obtain the eigenvalue spectrum without
explicitly solving  the differential equation.

The generic case in the context of  quantum mechanics was 
studied by Infeld  and Hull  who observed  that most of the 
Schr\"odinger and  Maxwell's equations as well as many other equations 
relevant to physics problems may be put in the  form\cite{Infeld,Wybourne,Barut}
\begin{align}
{{{d^2}{\cal R}}\over{ds}^2} + f(s) {\cal R}  = 0  \label{eq:calR00}
\end{align}
But there could be many potentials with different functional forms 
giving rise to the same eigenvalue, which reminds
the cases where one cannot hear the shape of the drum
uniquely. Symmetry  considerations have been exploited in finding such 
isospectral Hamiltonians.  One   approach is to find
the set of potentials in Hamiltonians with the similar eigenvalues.
Abraham and Moses attempted to search for exact solutions 
and energy eigenvalues  through symmetry arguments\cite{Abraham}.
 They extended the number of one dimensional  Schr\"odinger equations
having exact point eigenvalues by using Levitan-Gelfand equation.
This generated new exactly solvable potentials from the few
 potentials for which the Schr\"odinger equation is known to be 
solvable by adding or subtracting 
a finite number of eigenfunctions. This is the analogue of a well known 
result in scattering theory\cite{Bargmann,Gibbon}.
Other methods such as Darboux construction and  Marchenko equation 
were exploited by Luban and Pursey, thereby expanding the 
available techniques\cite{Pursey} .
Another development was the use of supersymmetry to  successfully 
combines two  essentially isospectral Hamiltonians into a single  
Schr\"odinger equation by introducing additional fermionic degrees 
of freedom\cite{Witten}. But as shown by Nieto the two supersymmetric 
Hamiltonians are related by a special case of
Darboux construction\cite{Nieto}.
There is an underlying algebraic structure arising out of the 
integrability condition known as shape invariance. This structure 
 can be identified with an associated Lie algebraic
 structure\cite{Balantekin}.  
These shape invariance algebra transforms the parameters such as 
strength and range. Shape invariance algebra, in general, 
are shown to be infinite dimensional.
 Balantekin has found the conditions under which the algebra 
becomes finite dimensional.
 Turbiner and Shifman  have also searched for exact solutions by 
symmetry arguments \cite{Turbiner}. They showed that the Hamiltonian 
of these quasi-exactly solvable potentials can be written in 
terms of symmetry generators, demonstrating the existence of 
a dynamical algebra in these potentials.
 The class of quasi-exactly solvable potentials have been  
further enhanced by De Souza and Boschi-Filho by working with an 
algebra isomorphic to $ sl(2)$
but using second order generators and writing the Hamiltonian as a
multilinear combination of these generators\cite{Dutra}.
The equation that governs  the isospectral deformations is in the form of
of a system of coupled Liouville equations\cite{Eleonesky}. If one 
eigenfunction and its corresponding
eigenvalue is exactly known then the group theoretic approach 
to intertwined Hamiltonians gives
new potentials\cite{Carinena}. Jafarizadeh and Fakhri\cite{Jafarizadeh} 
have shown that almost all solvable potentials with the nice property of 
parasupersymmetry and shape invariance are obtainable from the master 
functions  which are the special functions of mathematical
physics\cite{Levai,Jafarizadeh1}.

The shape invariant operators also form the generators of  a $gl(2,c)$ 
algebra.  The master functions determine the geometry of the two 
dimensional symmetric space.
The Casimir operator of $gl(2,c)$ algebra on these manifolds represents 
 the Hamiltonian operator of a charged particle moving in the presence 
 of some static background electric and magnetic field.
Two and three dimensional Hamiltonians have also been obtained 
with shape invariance property.  We discuss here all these developments
to  emphasize the special role  of the algebraic methods which are
being progressively  utilised in the  analysis of isospectral Hamiltonian and
related problems to get the results. 
 
Many aspects of the above developments originate from the observation of
Infeld  and Hull  that most of the Schr\"odinger equations
relevant to physics problems may be put in the 
form of equation (\ref{eq:calR00})\cite{Infeld,Wybourne,Barut}.
In this note we observe that a class of equations can be obtained from the 
above equation which all have the same Casimir operator.
A proper factorization of the equation corresponds to the generators
of $SU(1,1)$ or $SO(2,1)$ and the energy eigenvalue corresponds to the
Casimir invariant constructed out of the generators.
By a change of variables we can generate from
equation (\ref{eq:calR00}) 
 a class of equations whose $SU(1,1)$ or $SO(2,1)$   generators can be 
easily written down and the Casimir invariants can be obtained.
Though these resulting equations  are not Schr\"odinger equations,
except for a limited class of transformations, one can
calculate the eigenvalues corresponding to them in term of the parameters 
of the original  Schr\"odinger equation we started with. 

The generation of the spectrum associated with a second order differential
equation of the form
\begin{align}
 {{{d^2}{\cal R}}\over{ds}^2} + 
 ( {a\over{s^2}} + b s^2  +c ) \, {\cal R}  = 0, \label{eq:calR1} 
\end{align}
where $f(s)$ of eqn. (\ref{eq:calR00}) is given as
\begin{align}
f(s)  = {\frac{a}{s^2}} + b s^2  + c,
\end{align}
has been analyzed by several authors \cite{Wybourne,Miller}.
We indicate in brief the procedure to obtain the eigenvalues for 
such equations.
The Lie algebra of non-compact groups $SO(2,1) $ and $SU(1,1)$
can be realized in terms of a single variable by expressing the generators
\cite{Wyb150}
\begin{align}
{{\Gamma}_1} = { {\frac{{\p}^2}{{\p s}^2}} } + { \frac{\alpha}{s^2} }
               + { \frac{s^2}{16} } \\
{{\Gamma}_2} = - { \frac{i}{2 } } {\left( s { \frac{\p}{\p s} } + {
  \frac{1}{ 2} } \right ) } \\
{{\Gamma}_3} =   { \frac{ {\p}^2}{{\p s}^2} } + {\frac{\alpha}{s^2} }
               - {\frac{s^2}{16} },
\end{align}
so that the $\Gamma$'s satisfy the standard algebra
\begin{align}
 {[{{\Gamma}_1} ,{{\Gamma}_2} ]} = - i {{\Gamma}_3 } , \\
% \qquad
 {[{{\Gamma}_2} ,{{\Gamma}_3} ]} =  i {{\Gamma}_1 } , \\
%  \qquad
 {[{{\Gamma}_3} ,{{\Gamma}_1} ]} =  i {{\Gamma}_2 } . \label{eq:Gammacom}
\end{align}
The existence of the Casimir invariant for $su(1,1)$
\begin{align}
{{\Gamma}^2} = {{{\Gamma}_3}^2} -  {{{\Gamma}_1}^2} - {{{\Gamma}_2}^2}
\end{align}
is exploited to obtain the explicit form of $ {{\Gamma}_i}$'s.
The second order differential operator in equation (\ref{eq:calR00})
in terms of the $su(1,1)$ generators is now
\begin{align}
 { \frac{{\p}^2}{{\p s}^2}   } +       {\frac{a}{s^2} } + b s^2  +c % \right. \nonumber \\
%\left.  
=   { \left(  {\frac{1}{ 2}} + 8 b \right) } \, {{\Gamma}_1} +
 { \left( {\frac{1}{2}} - 8 b  \right) } {{\Gamma}_3}  + c,
\end{align}
and (\ref{eq:calR00}) becomes
\begin{align}
 \left[  { \left(  \frac{1}{ 2} + 8 b \right ) } \, {{\Gamma}_1} +
 {\left( \frac{1}{2} - 8 b  \right)} {{\Gamma}_3}  + c \right] \,  {\cal R} = 0  .
\end{align}
Next a transformation involving $ {e}^{-i \theta {\Gamma}_2 } $ can be
 performed
on $\cal R $ and the ${\Gamma}$'s . A choice of $ \theta $ such that
\begin{align}
{\tanh{\theta}} = - { \frac{  {\frac{1}{2} + 8 b}   }{ {\frac{1}{2}} - 8b }  },
\end{align}
will diagonalize the compact operator ${\Gamma}_3 $ and the discrete
eigenvalues may be obtained. The arguments of the standard
representation theory then leads to the result,
\begin{align}
4n + 2 + {\sqrt{1 - 4 a } } = { \frac{c}{\sqrt{ - b}} } ,\qquad
 n = 0, 1, 2, \dots   \label{eq:nabc}
\end{align}

 By a substitution
\begin{align}
s = {y^3},
\end{align}
 the equation(\ref{eq:calR1}) becomes 
\begin{align}
{ \frac{ {d^2}{\cal R} }{{dy}^2 }  } - 
{  {\frac{2}{ y} }  {\frac{d {\cal R}} {dy}}  }
 + 9  \left(   {\frac{a}{y^2} }  + by^{10} + cy^4 \right)  \, {\cal R}  = 0, \label{eq:calRy}
\end{align}
and the corresponding  $\Gamma$'s satisfying the commutation relations 
of Eq.({\ref{eq:Gammacom}}) are
\begin{align}
{{\hat{\Gamma}}_1} = { \frac{1}{9y^4} } {\frac{{\p}^2}{{\p y}^2}} 
  - { \frac{2}{9y^5} }  {\frac{\p}{\p y} }  +{\frac{\alpha}{y^6} }
               + {\frac{y^6}{16}} \\
{{\hat{\Gamma}}_2} = - {\frac {i} {2 }} { \left(  {\frac {y}{ 3}} \,
  {\frac{\p}{\p y} }
 + { \frac{1}{2} }\right) } \\
{{\hat{\Gamma}}_3} = {\frac{1}{9y^4}} {\frac{{\p}^2}{{\p y}^2}} 
  - { \frac{2}{9y^5}}  {\frac{\p}{\p y} }  +{\frac{\alpha}{y^6}}
               - {\frac{y^6}{16}}, 
\end{align}
which ultimately leads to a result similar to that of equation (\ref{eq:nabc})
for the Casimir.
Also if we let
\begin{align}
 {\cal R} = y { {\hat{\cal R}}(y)}, \label{eq:calRy} 
\end{align}
then we land up with the equation
\begin{align}
 {  \frac{{d^2}{\hat{\cal R}}}{{dy}^2}  } - { \frac{2}{y^2} } {\hat{\cal R}}  
 + 9 { \left( {\frac{ a }{y^2} } +
 b y^{10} + cy^4 \right ) }  {\hat {\cal R}}  = 0.  \label{eq:calRyhat}
\end{align}
Similarly, a substitution of the form
\begin{align}
s = { y^{\frac{2}{ 3}}},
\end{align}
and
\begin{align}
{\cal R}  = y^{-\frac{1}{ 6}} {\bar{\cal R}}
\end{align}
takes equation (\ref{eq:calR1}) to
\begin{align}
{  \frac{ {d^2}{\bar{\cal R}} }{{dy}^2} }
 + \left(  {\frac{a}{y^2} } - { \frac{5}{36 y^2} } + {\frac{4}{ 9}} b\, y^{\frac{2}{ 3}}
 + { {\frac{c}{y^{\frac{2}{3}}} }} \right) {\bar {\cal R}}  = 0,  \label{eq:calRyhat1}
\end{align}
and would have similar  Casimir.

In fact we can go from equation (\ref{eq:calR1}) to a class of equations
\begin{widetext}
\begin{align}
{  \frac{ {d^2}{u(x)} }{{dx}^2}   } 
 -  {    \left[   { \frac {( p^2 - 1)}{4x^2}  } %  \right. \nonumber \\
  %\left. 
+ p^2 {(   {  \frac {a}{ x^2}   } + 
b x^{(4p -2)}  + c x^{(2p - 2)}   )}  \right] } u(x) 
   = 0  \label{eq:gen}
\end{align}
\end{widetext}
that has similar Casimir when
\begin{align}
{\cal R} = {s^q u},
\end{align} 
with $s$ and $x$ are related by
\begin{align}
s = {x^p}.
\end{align}

Now, the original equation (\ref{eq:calR1}) has the energy eigenvalues
expressed in terms of the  constants $ a, b,$ and $c$ as in equation
 (\ref{eq:nabc}). 
Writing the equation (\ref{eq:calRyhat}) as an eigenvalue equation
by dividing  throughout $ {y^4}$, say, we have
\begin{align}
  {\frac{1}{y^4}}  { \frac{ {d^2}{\hat{\cal R}} }{{dy}^2} }
  - { \frac{2}{ y^{6}} }  {\hat{\cal R}}   % \nonumber \\
 + 9 \left(  {\frac{ a }{{y^{6}}} } + by^{6} + c \right) \,  {\hat {\cal R}}  
  = 0  \label{eq:calRyhat2}
\end{align}
with $ - 9\, c $ as the eigenvalue of the operator
\begin{align}
  {\cal{O}} = {\frac{1}{y^4} } {\frac{d^2}{{dy}^2} } 
  - { \frac{2}{ y^{6}} }   
 + 9 \,  \left(  {\frac{ a }{y^{6}}  } + b y^{6} \right).  
   \label{eq:calO}
\end{align}
Since we know the $c$ in terms of $ a, b, $ and $ n$,
the eigenvalues of the operator $\cal {O} $ are 
now obtained. In this way one can calculate the eigenvalues 
for the whole class of differential equations derivable from
our original equation  (\ref{eq:calR1}) by appropriate change of 
variables. This is the central result of our analysis.
The above considerations can be used for similar problems
where the algebraic method of calculating the eigenvalues
is possible such as the three dimensional harmonic oscillator.

We may also interprete (\ref{eq:calRyhat}) as Schro\"odinger equation with
a potential 
\begin{align}
{ \frac{2}{ y^2} } + 9 \left( { \frac{ a}{y^2} } + b \, y^{10} + c \,y^4\right),
\end{align}
 which gives a zero eigenvalue. This would also lead to a hierachy 
of equations with zero eigenvlues if we substitute the appropriate 
eigenvalues and eigenfunctions for the original Schro\"odinger 
equation (\ref{eq:calR00}), providing algebraic relationships 
between different eigenfunctions.

The solutions to the above differential equations are easy to obtain as 
the basic equation  (\ref{eq:calR00}) can be solved by Frobenius series
expansion method or by the formal Lie-algebraic approach\cite{Miller}. The solutions to the 
other equations are obtained by  just multiplying by the appropriate functions. \\

Physical considerations in many cases forces one to consider
potentials which may not be analytically solvable. For example Maxwell
had shown that in the dynamical theory of gases the problem can be
solved exactly for transport coefficients in a gas if the
intermolecular force varies as the inverse fifth power of the 
intermolecular distance\cite{Chandrasekhar}. The Lenard-Jones potential has the form
$ V_{LJ} = 4\varepsilon \left[ \left(\frac{\sigma}{r}\right)^{12} -
  \left(\frac{\sigma}{r}\right)^{6} \right] = \varepsilon \left[
  \left(\frac{r_{m}}{r}\right)^{12} -
  2\left(\frac{r_{m}}{r}\right)^{6} \right]$,
where $\varepsilon $ is the depth of the potential well, $\sigma$  is the finite distance
at which the inter-particle potential is zero, $ r$ is the distance
between the particles, and $ r_{m }$ is the distance at which the potential
reaches its minimum. Such type of phenomenological potentials 
give much insight to explain many physical properties of a system.
Algebraic methods may be helpful in attacking such cases.

Since a similar analysis can be made for the continuum  eigenvalues, 
it is natural to expect that these equations will have some symmetries
related to
B\"acklund symmetries. However, we have not yet been able to 
establish the connection.

The aim of this note is to draw attention to the fact that by considering
the algebraic method  of group theory, starting with a Schr\"odinger
type of second order equation with ${  { \frac{a}{r^2} } + b r^2 } $ form of
 potential,
a class of  equations can be obtained whose eigenvalues can be calculated
 with the  help of the Casimir operator. This is achieved without explicitly
solving these equations. Since   the underlying symmetry algebra of our
 differential equations is $su(1,1)$, the solutions are expected  to be  
related to the special 
functions of mathematical physics.
The normalization of the solutions and related problems are 
left for future study.

\begin{acknowledgments}
I am grateful to  David A. Vogan for an enlightening
discussion and for the hospitality at the Massachussetts Institute of Technology.
\end{acknowledgments}

\end{document}